\documentstyle[prl,aps,epsf]{revtex}
\begin{document}
\draft
\title{Chaos-Assisted Light Squeezing}
\author{Kirill N. Alekseev$^{1,2}$\cite{email1},  
and Jan Pe\v{r}ina$^{1}$\cite{email2}}
\address{$^1$Laboratory of Quantum Optics,
 Palack\'{y} University, T\v{r}\'{\i}da Svobody 26, 771 46 Olomouc,
 Czech Republic \\
$^2$Kirensky Institute of Physics,
 Russian Academy of Sciences, Krasnoyarsk 660036, Russia }
\maketitle
\begin{abstract}
We investigate theoretically the dynamics of squeezed state generation
in the nonlinear systems possessing the transition from regular to chaotic
dynamics in the limit of a large number of photons. As an example, the
model of kicked Kerr oscillator is considered. 
We show the direct
correlation of the degree of squeezing and the value of local Lyapunov
instability rate in corresponding classical chaotic system.
\end{abstract}
%\pacs{
\par
In recent years the problem of generation of the squeezed states of
electromagnetic field has attracted a great attention of both the
theoreticians
and experimentalists \cite{1,2,3,4}. As a rule, in the experiments on
squeezing the quantum fluctuations are small compared to the mean value of
the field modes involved into a nonlinear interaction \cite{3,4}. Already
for a rather long time it was mentioned that a light squeezing can be
increased near the bifurcation points between different dynamical regimes
\cite{5,6,7,8,9}. The reason for such enhancing consists in a strong
diverging of one of the field quadrature component and corresponding
squeezing of another one at the vicinity of instability threshold \cite{5,9}.
Among with well studied parametric media \cite{5,9}, the increase of squeezing
was predicted for the interaction of the field with 2-level atoms inside a
high-$Q$ cavity at the dynamical regime corresponding to the separatrix
\cite{6,7,8}. However, all papers devoted to the study of the enhanced
squeezing due to the appearance of dynamical instabilities in optical systems
dealt with the {\it integrable or near integrable systems 
with regular dynamics}
\cite{remark1}.
\par
In this paper, we consider the squeezed light generation by quantum
nonintegrable optical system obeying the transition from regular to chaotic
dynamics in the classical limit. As for other problems of quantum 
chaos \cite{11,12}, we consider
the semiclassical limit when a great number of quantum levels $N\gg 1$ are
involved into the dynamics. Our prime goal is to demonstrate the sufficient
increase of the degree of light squeezing at transition to quantum chaos.  
We also found direct correlation between the
degree of local instability in corresponding classical system and the
degree of light squeezing.
\par
In spite of our consideration is valid for any nondissipative quantum system
with one and half degrees of freedom, we will demonstrate our main results on
the enhanced squeezing for some particular model of the nonintegrable optical
system -- the nonlinear oscillator periodically forced by the classical
field. In the interaction picture the Hamiltonian has the form
($\hbar\equiv 1$)
% Eq(1)----------------
\begin{equation}
\label{1}
H=\Delta b^{+}b + \frac{\lambda}{2} b^{+2}b^2 +
\varepsilon  N^{1/2} (b + b^+) F(t),
\end{equation}
%----------------------
where Bose operators $b$ and $b^+$ ($[b,b^+]=1$) describe a single-mode
of the  quantum field and $\lambda$ is proportional to third-order nonlinear
susceptibility of a nonlinear medium. The last term in (\ref{1}) corresponds
to a coupling of the oscillator with external classical periodically
modulated field containing a large number of photons $N\gg 1$ ($\varepsilon
$ is a coupling constant, $F(t)$ is a periodic function of time and $\Delta$
is a detuning of the mode frequency from the carrier frequency of the
external field).
\par
The Hamiltonian (\ref{1}) describes, for example, a high-$Q$ cavity filled
by a medium with Kerr nonlinearity and excited by an external laser field
\cite{kukl}.
This effective Hamiltonian may also govern the interaction of a laser field
with a high-density exciton system in a semiconductor \cite{14}. 
The differen
variants of discussed model are very popular in both quantum optics \cite{15}
and quantum chaos \cite{12,16',16,17,18} studies.
\par
To investigate the nonlinear dynamics of the model and the dynamics of
squeezing, it is natural to use the cumulant expansion technique \cite{19}
as a variant of the general $1/N$-expansion method \cite{20}. Recently it
was realized that both methods could be applied not only to the integrable
\cite{7,19,20} but also to the nonintegrable quantum systems \cite{16,21,22}.
In this paper, we use well-adopted for the problems of quantum optics th
$1/N$-expansion method suggested in \cite{7}.
\par
Introduce new normalized operators for annihilation and creation of photons
as $a=b/N^{1/2}$, $a^+=b^+/N^{1/2}$ with commutation relation $[a,a^+]=
1/N$. In the classical limit ($N\rightarrow\infty$), one has two commuting
$c$-numbers. Now the Hamiltonian (\ref{1}) can be rewritten as $H=N H_N$, 
where
$H_N$ has the same form as (\ref{1}) with an account of replacement
$b\rightarrow a$, $b^+\rightarrow a^+$ and $\lambda\rightarrow g\equiv
\lambda N$. It may be shown that the dependence $g\simeq N$ correctly
performs  the time scale of energy oscillation for Kerr nonlinearity in
the classical limit \cite{23}.
\par
Let initially the field is in the coherent state $\mid\alpha\rangle=
\exp(N\alpha a^+ -N\alpha^* a) \mid 0\rangle$ corresponding to the mean
photon number $\simeq N$. From the Heisenberg equations for
$a$, $a^2$ and their Hermitian conjugated equations, and adopting the normal
operator ordering, we have the following equations of motion for averages
over coherent states
% Eq (2) ------------------------------------
\begin{equation}
\label{2}
i \frac{d}{d t}\langle\alpha\rangle  =
\langle V\rangle, \quad
i \frac{d}{d t}\langle\left(\delta\alpha\right)^2\rangle =
2\langle V\delta\alpha\rangle+
\langle W\rangle, \quad
i \frac{d}{d t}\langle\delta\alpha^*\delta\alpha\rangle =
-\langle V^*\delta\alpha\rangle+
\langle\delta\alpha^* V\rangle,
\end{equation}
%--------------------------------------------
where $V=\frac{\partial H_N}{\partial a^+}$,
 $W=\frac{1}{N} \frac{\partial V}{\partial a^+}$,
 $\langle\left(\delta\alpha\right)^2\rangle\equiv
\langle a^2\rangle - \langle a\rangle^2$,
 $\langle\delta\alpha^*\delta\alpha\rangle\equiv
\langle a^+ a\rangle - \langle a^+\rangle \langle a\rangle$.
However, the set of equations (\ref{2}) is not closed and actually is
equivalent to the infinite  dynamical hierarchy system for moments and
cumulants \cite{7,19}. To truncate it up to the cumulants of the second
order, we made the substitution $a\rightarrow\langle\alpha\rangle +
\delta\alpha$, where at least initially the mean $\langle\alpha\rangle\simeq
1$ and the quantum correction $\mid\delta\alpha(t=0)\mid\simeq N^{-1/2}\ll 1$.
Using the Taylor expansion of the functions $V$ and $W$, we have, in the first
order of $1/N$ and after some algebra, the following self-consistent system
of equations for the mean value and the second order cumulants \cite{remark2}
\begin{mathletters}
\label{3}
% Eq (3a) -------------------------------------------
\begin{equation}
i\dot{z}=\langle V\rangle_z + q,
\label{3a}
\end{equation}
%--------------------------------------------------
% Eq (3b)--------------------------------------------
\begin{equation}
i\dot{C}=2\left(\frac{\partial V}{\partial\alpha}\right)_z C +
2\left(\frac{\partial V}{\partial\alpha^*}\right)_z B,
\label{3b}
\end{equation}
%---------------------------------------------------
% Eq (3c)-----------------------------------------
\begin{equation}
i\dot{B}=-\left(\frac{\partial V^*}{\partial\alpha}\right)_z C +
\left(\frac{\partial V}{\partial\alpha^*}\right)_z C^*,
\label{3c}
\end{equation}
%-------------------------------------------------
\end{mathletters}
where $B\equiv \langle\delta\alpha^*\delta\alpha\rangle+\frac{1}{2N}$,
$C\equiv\langle\left(\delta\alpha\right)^2\rangle$,
$z\equiv\langle\alpha\rangle$ and subscript $z$ means that
the values of $V$ and its derivatives
are calculated at mean value $z$. Involved to the equation (\ref{3a})
the small quantum correction $q\simeq 1/N$ has the form of the second
differential of  $V$ as follows
% Eq without number---------------
$$
q=\frac{1}{2} d^2 V\mid_z =
\frac{1}{2} \left(\frac{\partial^2 V}{\partial\alpha^2}\right)_z C +
\frac{1}{2} \left(\frac{\partial^2 V}{\partial\alpha^{*2}}\right)_z C^* +
\left(\frac{\partial^2 V}{\partial\alpha^{*}\partial\alpha}\right)_z
\left(B-\frac{1}{2N}\right).
$$
%----------------------------------------------------------
The initial conditions for the system (\ref{3}) are $B(0)=\frac{1}{2N}$,
$C(0)=0$ and some arbitrary $z(0)$ which is of the order of unity.
\par
It is easy to see that the equations of motion (\ref{3b}) and (\ref{3c})
for the cumulants can be obtained from the classical equations by
 linearization
near $z$ (substitution $z\rightarrow z+\Delta\alpha$, $\mid\delta\alpha\mid
\ll\mid z\mid$), if one writes the dynamical equations for the variables
$(\Delta\alpha)^2$ and $\mid\Delta\alpha\mid^2$. But there still exists the
principal difference between the linearization of the classical motion
equations and the equations for quantum cumulants (\ref{3}): {\it
it is impossible
to obtain the initial conditions for $C$ and $B$ from only initial conditions
for linearized classical equations}.
\par
Define the general field quadrature as $X_\theta=a\exp(-i\theta)+
a^+\exp(i\theta)$,
where $\theta$ is a local oscillator phase. A state is said to be squeezed
if there is some angle $\theta$ for which
the variance of $X_\theta$ is smaller then the variance for a coherent
state or the vacuum
\cite{1,2}. Minimizing the variance of $X_\theta$ over $\theta$, one can
determine the minimum half-axis  of the quantum noise ellipse. Then, the
condition of the {\it principal squeezing} is
$S\equiv 1+2 N (\langle\mid\delta\alpha\mid^2\rangle-
\mid\langle(\delta\alpha)^2\rangle\mid)=
2 N (B-\mid C\mid) < 1$ \cite{2}.
The determination of the principal squeezing $S$ is very useful because it 
gives {\it the maximal squeezing measurable by the homodyne
detection} \cite{2}.
\par
Let us now compare the dynamics of the principal squeezing for regular
and chaotic motion.
 Initially Gaussian wave packet  spreads when it  propagates through 
 nonlinear medium. But it still exists the time interval of the 
 well-defined quantum-classical correspondence during which the wave 
 packet center follows path in the phase space governed by semiclassical 
 equations of motion (\ref{3a}). Moreover, because our equations of motion 
 for cumulants (\ref{3b}) and  (\ref{3c}), in fact,
coincide with the equations arising in the definition of the maximal Lyapunov
exponent \cite{11,12},  we can apply simple physical arguments on strong 
deformation of the classical phase volume at chaos for prediction of the 
strong squeezing of the noise ellipse at quantum chaos in the semiclassical 
limit. For chaotic dynamics, the  distance in the phase space between two 
initially very closed trajectories $D$ grows exponentially with time 
$D(t)\simeq\exp(\lambda t)$, where $\lambda$ is the maximal Lyapunov exponent.
 Due to the presence
of local strong (exponential) local instability inherent for underlying 
classical
chaotic dynamics, a quantum noise ellipse may be strongly stretched in one
direction and squeezed in another direction. As a result, the value of the 
principal squeezing in average exponentially decreases in time at chaotic 
dynamics. 
The stretching and squeezing of noise ellipse at quantum chaos is much 
stronger than for the case of regular and stable dynamics, when the distance 
between two initially closed trajectories in phase space increases in time 
power-wise
resulting in only power-wise decreasing of the principle squeezing
in time.
\par
In order to illustrate this general picture, we return to Hamiltonian 
(\ref{1})
and choose  the form of $F(t)$ as a periodic sequence of kicks: $F(t)=
\sum_{n=-\infty}^{\infty}\delta(t-nT)$, where $\delta(t)$ is
the Dirac $\delta$-function. In an experiment, a sequence of
short light pulses can be generated by a mode locked laser.
Now, by means of standard technique \cite{12}, we obtain from 
the differential equations (\ref{3})
the coupled maps for the mean value and for the second order
cumulants. In this paper, we present only the form of the resulting
map for the mean value \cite{remark3,26'}
% Eq (4)------------------------------------------------
\begin{equation}
\label{4}
z_{n+1}=e^{-i T \left(\Delta+g\mid z_n-i\varepsilon\mid^2\right)}
(z_n-i\varepsilon)+q_n(C_n,B_n,z_n),
\end{equation}
%---------------------------------------------------------
where the subscript $n$ indicates the value of a function just before
the action of the $n$-th kick. The use of the maps instead of the differential
equations sufficiently simplifies the computation and reduces numerical
errors.
\par
In the classical limit $N\rightarrow\infty$ and $q\rightarrow 0$,
the map (\ref{4}) possesses  the transition from regular to chaotic dynamics.
Fig. 1 illustrates the behavior of the classical map: the phase portrait and
the time-dependence of intensity $\mid z_n\mid^2$ for regular (Fig. 1a,b) and
chaotic (Fig. 1c,d) dynamics.
\par
Define the distance in the phase space between two
initially very closed trajectories as $D(t)=[(\delta x)^2+(\delta y)^2]^{1/2}$,
where $x=Re z$ and $y=Im z$.
The Fig. 2 shows the time-dependence of the logarithms 
of the principal squeezing $S$ (Fig. 2a), the normalized distance 
between two initially very
closed trajectories $d=N D$ (Fig. 2b), and the normalized value of quantum
correction $Q=N q$ (Fig. 2c) for large but finite number of quanta $N=10^9$.
In this figure, curve 1 corresponds to the regular
motion, and curves 2 and 3 - to chaotic motion with slightly different
values of the
Lyapunov exponent. As it is evident from this figure, the most strong local
instability determines the highest degree of squeezing. It should be noticed
that the difference in the magnitude of principal squeezing for chaotic and
regular motion may achieve of several orders during only  several
kicks.
\par
 As follows from (\ref{3}), for chaos, both cumulants $B$
and $C$ increase also exponentially in average, resulting in an exponential
growth of the quantum correction $q$.
All our analysis is valid, if and only if the values of the quantum
correction and of the second order cumulants are much less than the mean
values: $q, \mid C\mid, B\ll\mid z\mid\simeq 1$. Under the conditions of
chaos, this means that the
equations (\ref{3}) are correct only during the time interval $t\ll t^*
\simeq\lambda^{-1}\ln N$. 
The time interval $t^*$ coincides with time scale determining the well-defined
quantum-classical correspondence in chaotic systems \cite{16'}.
Nowadays the time scale $t^*$, during which classical chaos is revealed
as a quantum transient, is established for the different models of 
quantum chaos \cite{12,16,22}.
Moreover, only during this time interval it is possible to define the
Lyapunov exponent for a quantum
system \cite{26}. Our numerical experiments on the model of kicked Kerr
oscillator demonstrate, that for the chaotic motion at $g T=7$,
$\Delta/g=1$ and $\varepsilon=0.1$, the value of the principal squeezing $S$
 for $N=10^7$ is practically indistinguishable from its value 
 for $N=\infty$ up
to the six kicks. The increase of the number of photons $N$ results in the
corresponding increase of the time interval for applicability of our
description of squeezed dynamics. On other hand, the number of photons
$N\ge 10^7$ initially pumped to the system is quite realistic for
contemporary experiments on light squeezing \cite{1,2,3,4}.
\par
In summary, we have showed that among with very narrow class of the 

integrable systems near threshold of an instability, there is another 
wide class of potentially effective  systems for the  enhanced light 
squeezing - the systems with quantum chaos operating during the time scale
of the well-defined quantum-classical correspondence.
\par
We would like to thank Evgeny Bulgakov, Claude Fabre, Zden\v{e}k Hradil, 
Feodor Kusmartsev,
Anton\'{\i}n Luk\v{s}, Vlasta Pe\v{r}inov\'{a}
for discussions. We also thank
Peter Milonni, Boris Chirikov and anonymous referee for bringing our 
attention to refs \cite{22,10',26'}. 
This work was partially supported by Russian Fund for Basic Research
(96-02-16564-a), INTAS (94-2058), Czech Grant Agency (202/96/0421),
and Palack\'{y} University.
\newpage
\vspace{2cm}
\epsfxsize=10cm
\hspace{2cm}
\epsfbox{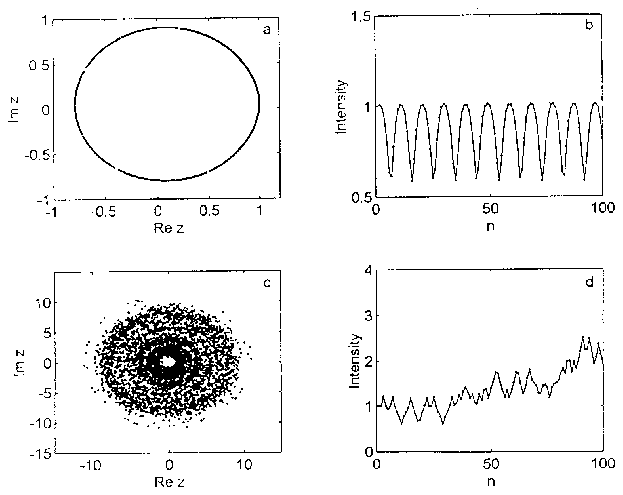}
\begin{figure}
\caption{The nonlinear dynamics of kicked classical  nonlinear oscillator:
phase portrait
and time-dependence of the intensity $\mid z_n\mid^2$ for the regular behavior
at $g T=3$ (a, b) and for the chaotic motion at $g T=10$
(c, d). Initial condition is $z_0=1$ and $\varepsilon=0.1$, $\Delta/g=1$.}
\label{fig1}
\end{figure}
\vspace{1.5cm}
\epsfxsize=10cm
\hspace{2cm}
\epsfbox{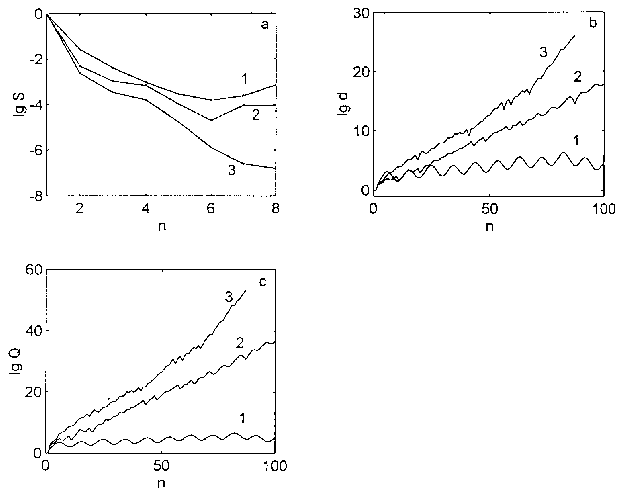}
\begin{figure}
\caption{The time-dependence of the principal squeezing (a), of the distance
between two initially
closed trajectories (b), and of the quantum correction (c). Curve 1
corresponds to
the regular dynamics at $g T=3$, curve 2 -- to the mild chaos ($g T=7$) and
curve 3 -- to the hard chaos ($g T=10$). The average photon number is $N=10^9$,
initial condition and other parameters are the same as in Fig. 1. }
\label{fig2}
\end{figure}


\begin{references}
\bibitem{9}
Very nice discussion of large squeezing near bifurcation points is
presented in ref. \cite{4}, sec. 3.3.1., see also references cited therein.
\bibitem{remark1}

The only exceptions are the papers \cite{8,10'}. In \cite{8}
the attempt was made to study the light squeezing in some particular
system -- generalized Tavis-Cummings model, which is chaotic in the classical
limit. In contrast, the results of the present paper are valid for an
arbitrary single-mode quantum system with the time-dependent Hamiltonian.
Strong squeezing of a wave-packet during time scale of well-defined 
quantum-classical correspondence also has been briefly discussed in \cite{10'}.
\bibitem{remark2}
The same self-consistent system can be obtained obeying another truncation
schemes. For example, if one makes the decoupling corresponding to
the hypothesis that during evolution a quantum state appears which is the
superposition of the coherent signal and the quantum noise \cite{2},
then the resulting system of equations for the forced nonlinear
oscillator \cite{17} coincides with our system of equations (\ref{3})
up to the terms of $O(1/N^2)$.
\bibitem{remark3}
The formulas describing coupled maps for the cumulants $B$ and $C$ and the
quantum correction $q$ are too long, therefore we will present them
together
with the detailed description of computation in our future full-size paper.
\end{references}
\end{document}